\documentclass[aps,nofootinbib]{revtex4-1}
\usepackage{graphicx,subfigure}
\usepackage{amsmath,amssymb}

\def\beq{\begin{equation}}
\def\eeq{\end{equation}}
\def\beqr{\begin{eqnarray}}
\def\eeqr{\end{eqnarray}}
\def\bdpm{\begin{displaymath}}
\def\edpm{\end{displaymath}}

\newcommand{\Zp}{Z^\prime}
\newcommand{\Zpp}{Z^{\prime\prime}}
\newcommand{\gVps}{g_V^{\prime 2}}
\newcommand{\gAps}{g_A^{\prime 2}}

\begin{document}

\title{
LHC Phenomenology of $Z^\prime$ and $Z^{\prime\prime}$ bosons 
in the $SU(4)_L \times U(1)_X$ little Higgs model 
}

\date{\today}

\author{Kang Young Lee}
\email{kylee.phys@gnu.ac.kr}

\affiliation{ 
Department of Physics Education \&
Research Institute of Natural Science,
Gyeongsang National University, Jinju 660-701, Korea
}

\author{Soo-hyeon Nam}
\email{glvnsh@gmail.com}

\affiliation{ 
Supercomputing Center, 
KISTI, Daejeon 305-806, Korea
}

\begin{abstract}
We examine direct limits on masses of the extra neutral gauge bosons
in the $SU(4)_L \times U(1)_X$ model with a little Higgs mechanism
confronted with the LHC data, 
especially by embedding anomaly-free set of fermions.
There exist two extra neutral gauge bosons, calling $\Zp$ and $\Zpp$,
in this model.
The lower exclusion limit of the mass of the lighter extra neutral gauge boson
is about 3 TeV while that of the heavier one 5 TeV.
For comparison, we examine the mass limit of $\Zp_3$ boson in the $SU(3)_L \times U(1)_X$ model as well,
and discuss the implication of our result in the $SU(4)_L \times U(1)_X$ model with a standard Higgs mechanism.
We also discuss the discovery potential of $\Zp$ and $\Zpp$
at the future LHC with the center-of-momentum energy of 14 TeV.
Our results can be applicable to the models with regular Higgs mechanism
if the same type of fermion family is assigned.
\end{abstract}

\maketitle

\section{Introduction}

 One of the major goals of the Large Hadron Collider (LHC) experiment is the search for heavy resonances.
Since a heavy resonance indicates a new heavy particle, its discovery is a clear evidence of the new physics (NP) 
beyond the standard model (SM). 
Especially the existence of an extra neutral gauge boson $\Zp$ is a common feature of many NP models.   
Therefore, the LHC has been pushing ahead on the search for neutral heavy resonances in various NP models. 
Recently CMS and ATLAS collaborations have reported the search results for the $\Zp$ existed in some NP models
through the dilepton channels at $\sqrt{s} = $7 and 8 TeV with the data up to 20 fb$^{-1}$, 
and also through $t\bar{t}$, $\tau^+\tau^-$, and dijet channels with similar or less integrated luminosities
\cite{CMS13jet, CMS12top, CMS12lepton, CMS12tau, CMS13lepton, ATLAS12lepton, ATLAS13top, ATLAS13tau,
CMS13jet0, CMS13lepton0, ATLAS12jet0, ATLAS13lepton0, ATLAS13top0, ATLAS13tau0}. 
Although we have not seen a new resonance yet so far, in future experiments,
we are still expecting to observe one or more neutral heavy resonance
due to various theoretical motivations \cite{Langacker08}. 
In this paper, 
we consider the NP model  with the electroweak $SU(4)_L \times U(1)_X$ gauge group
where there are two extra neutral heavy bosons,
and probe the search possibility of those heavy bosons at the LHC, especially by embedding anomaly-free set of fermions.

 The $SU(4)_L\times U(1)_X$ extension of the electroweak gauge symmetry $SU(2)_L \times U(1)_X$
of the SM has been widely studied by various authors 
due to its several distinctive features \cite{Fayyazuddin84, Voloshin88, Pleitez93, Ponce07, Kaplan03}.
One of the most interesting feature is that the gauged $SU(4)_L\times U(1)_X$ group including both quarks and leptons 
can provide an answer to the question why we only observe three families of fermions in nature, 
in a sense that anomaly cancellation is achieved when $N_f = N_c = 3$ 
where $N_f (N_c)$ is the number of families (colors) \cite{Pleitez93}. 
Also, the $SU(4)_L \times U(1)_X$ gauge structure has been recently considered 
in order to implement the little Higgs mechanism by Kaplan and Schmaltz (K-S) 
as an alternative solution to the hierarchy and fine-tuning issues \cite{Kaplan03}.
The noble feature of the models with the little Higgs mechanism is that the Higgs mass is protected by a global symmetry 
which is spontaneously broken 
and so the one-loop quadratic divergences to the Higgs mass are canceled by particles of the \textsl{same} spin; 
i.e., a new fermion cancels a quadratic divergence from a SM fermion.
However, the K-S model has a set of anomalous fermions with a simple family-universal embedding 
which requires additional fermion multiplets at the scale $\Lambda$ due to nonvanishing quadratic divergences 
from light fermions and gauge anomalies.  
Alternatively, we consider the modified K-S model by embedding anomaly-free set of fermions as proposed in Ref. \cite{Kong03} 
so that one-loop quadratic divergences for the Higgs mass are canceled for all fermion flavors
as a continuation of our previous work \cite{Nam10}. 

 Due to the different gauge group structure and the multiple breaking of global symmetry by separate scalar fields,
the neutral currents in this model are fundamentally different from those in other types 
of the little Higgs models (LHMs) \cite{Schmaltz05}.
In this model with anomaly-free set of fermions,
there are two extra massive neutral gauge bosons, $\Zp$ and $\Zpp$, which couple to ordinary SM fermions,
while other extra gauge bosons only couple to new heavy fermions.
Therefore, in this paper, we mainly focus on the LHC phenomenology of $\Zp$ and $\Zpp$ bosons.
Similar neutral gauge bosons appear in other $SU(4)_L \times U(1)_X$ models with regular Higgs mechanism \cite{Ponce07},
and their low-energy phenomenology is also similar if the fermion family is assigned to be the same \cite{Nam10}. 
As a result, the mass bounds and the LHC discovery potentials of the $\Zp$ and $\Zpp$ bosons obtained in this paper 
can be applicable to the models with regular Higgs mechanism.  
For comparison, we also consider the LHC phenomenology of the $\Zp$ boson 
appeared in the LHM with the $SU(3)_L \times U(1)_X$ gauge symmetry 
in which the anomaly-free set of fermions are embedded as well \cite{Schmaltz04, Han06}.
In order to have a proper Higgs mass, the $SU(3)$ LHM requires so-called $\mu$-term that manifestly breaks the global $SU(3)^2$ symmetry and gives tree-level masses to scalar particles including Higgs, 
which ruins the original motivation of the LHM to acquire Higgs mass spontaneously \cite{Kaplan03}.
Nonetheless, the LHC phenomenology of the $\Zp$ boson in the $SU(3)$ LHM is also similar to 
that of the $\Zp$ boson appeared in the usual 3-3-1 models with regular Higgs mechanism \cite{Coutinho13}, 
so we take into account the $SU(3)$ LHM as a benchmark model. 

 This paper is organized as follows. The generic structure of the model is discussed in the next section.  
We present the decay properties of the $\Zp$ and $\Zpp$ bosons in the section III, 
and obtain the lower bounds on their masses from the present LHC data in the section IV.
In the section V, we study the discovery potentials of the $\Zp$ and $\Zpp$ bosons at the next stage of the LHC 
with 14 TeV center-of-mass (CM) energy.
Finally, we conclude in the section VI.

\section{$SU(4)_L \times U(1)_X$ model with little Higgs}

The LHMs adopts the early idea that Higgs can be considered as a Nambu Goldstone boson 
from global symmetry breaking at some higher scale $\Lambda \sim 4\pi f$ \cite{Dimopoulos82} 
and acquires a mass radiatively through symmetry breaking at the electroweak scale $v$ by collective breaking \cite{Arkani01}.
Especially, the $SU(4)_L \times U(1)_X$ LHM is characterized by the scalar sector based on the non-linear sigma model 
describing $[SU(4)/SU(3)]^4$ global symmetry breaking with the diagonal $SU(4)$ subgroup gauged 
and four non-linear sigma model field $\Phi_i$ parameterized as
\beq
\Phi_{1}=e^{+i {\cal H}_u \frac{f_{2}}{f_{1}}}
        \left( \begin{array}{l} 0  \\ 0 \\ f_{1} \\0 \end{array} \right) ,\quad
\Phi_{2}=e^{-i {\cal H}_u \frac{f_{1}}{f_{2}}}
        \left( \begin{array}{l} 0  \\ 0 \\ f_{2} \\0 \end{array} \right) ,
        \nonumber
\eeq
\beq \label{Eq:Scalarfield}
\Phi_{3}=e^{+i {\cal H}_d \frac{f_{4}}{f_{3}}}
        \left( \begin{array}{l} 0  \\ 0 \\ 0 \\f_{3} \end{array} \right) ,\quad
\Phi_{4}=e^{-i {\cal H}_d \frac{f_{3}}{f_{4}}}
        \left( \begin{array}{l} 0  \\ 0 \\ 0 \\f_{4} \end{array} \right) ,
\eeq
where
\beq \label{Eq:Higgs}
  {\cal H}_u =\left[\Pi_u +
    \left( \begin{array}{ccc}
           \begin{array}{cc} 0 & 0 \\ 0 & 0 \end{array}
             & h_u & \begin{array}{c} 0 \\ 0 \end{array}   \\
            h_u^\dagger & 0 &  0 \\
           \begin{array}{cc}  0 &  0 \end{array} & 0 & 0 \\
           \end{array} \right)\right]\Big{/}f_{12} ,
\quad
  {\cal H}_d =\left[\Pi_d +
    \left( \begin{array}{ccc}
      \begin{array}{cc} 0 & 0 \\ 0 & 0 \end{array}
      &  \begin{array}{c}  0 \\  0 \end{array}  &  h_d \\
        \begin{array}{cc}  0 &  0 \end{array}  & 0 &  0 \\
      h_d^\dagger &  0 & 0 \\
    \end{array} \right)\right]\Big{/}f_{34} ,
\eeq
with $f_{ij} = \sqrt{f^2_{i} + f^2_{j}}$.  Here we only show the two complex Higgs doublets $h_{u,d}$ and discard the other singlets 
in $\Pi_q\ (q=u, d)$ whose contributions to fermion and gauge boson masses are negligible.  
The two doublet Higgs fields $h_{u,d}$ shown in Eq. (\ref{Eq:Higgs}) are of the following form
\beq \label{Eq:Higgsfield}
h_u = \frac{1}{\sqrt{2}}{H_u^0 \choose H_u^-} , \quad  h_d = \frac{1}{\sqrt{2}}{H_d^+ \choose H_d^0},
\eeq
where the neutral components of the two Higgs fields develop vacuum expectation values (VEVs) such that
\beq \label{Eq:Higgsvev}
\langle H_u^0 \rangle =  v_u  , \quad \langle H_d^0 \rangle =  v_d  .
\eeq
The Higgs vacua introduced above give masses to the SM fermions after the electroweak symmetry breaking (EWSB). 

 In this model, the $SU(4)$ breaking is not aligned and only the gauged $SU(2)$ is linearly realized.
The SM gauge group $SU(2)_L\times U(1)_Y$ can be embedded into the theory with an additional $U(1)_X$ group.
Since this LHM has a gauged $SU(4)_L$, the SM doublets must be expanded to $SU(4)_L$ quadruplets.
The extra fermions in the quadruplets should cancel the quadratic divergence from the SM fermion, especially from the top quark.  
Taking into account this requirement, we embed the SM doublet ($t,b$) into the following type of
$SU(4)_L$ quadruplet as 
\beq \label{Eq:fermion}
\psi_L = (t, b, T, B)_L^T,
\eeq 
so that the duplicated extra heavy fermions $T$ and $B$ remove the quadratic divergences due to their SM fermion partners $t$ and $b$, respectively, as discussed in Ref. \cite{Nam10}.  
 After the EWSB, among the 15 gauge fields $A_\alpha^\mu$ associated with $SU(4)_L$, 
the three neutral gauge bosons $A^3$, $A^8$ and $A^{15}$ mixed with the $U(1)_X$ gauge boson $A^X$ 
are associated with a $4\times 4$ nondiagonal mass matrix. 
After the mass matrix is diagonalized, a zero eigenvalue corresponds to the photon $A$, 
and the three physical neutral gauge bosons $Z$, $\Zp$ and $\Zpp$ get the masses as follows \cite{Nam10}
\beq \label{Eq:Zmass}
M^2_Z = \frac{g^2v^2}{4c_W^2}\left(1 - \frac{t_W^4}{4}\frac{v^2}{f^2}\right), \quad
M^2_{\Zp} = (g^2+g_X^2)f^2 - M^2_{Z}, \quad
M^2_{\Zpp} = \frac{1}{2}g^2f^2 ,
\eeq
where $g$ and $g_X$ are the couplings of the $SU(4)_L$ and $U(1)_X$ gauge groups, respectively,
and where the Weinberg mixing angle $\theta_W$ is 
\beq
c_W \equiv \cos{\theta_W} = \sqrt{\frac{g^2+g_X^2}{g^2+2g_X^2}} , 
\eeq
and the VEV $v$ is given by $v^2 = v_1^2+v_2^2$ with
\beq
v_1^2 = v_u^2 - \dfrac{v_u^4}{3f^2}
  \left(\dfrac{f_{2}^2}{f_{1}^2}+ \dfrac{f_{1}^2}{f_{2}^2}-1\right), \quad
v_2^2 = v_d^2 - \dfrac{v_d^4}{3f^2}
  \left(\dfrac{f_{4}^2}{f_{3}^2}+ \dfrac{f_{3}^2}{f_{4}^2}-1\right). 
\eeq
Note that the following simplifying assumption $f_{12} = f_{34} = f$ is used as done in Ref. \cite{Kaplan03, Csaki03}.  
Under this assumption, $\Zpp$ does not mix with $Z$ or $\Zp$ but still couples to the ordinary SM fermions.
There are also extra flavor-changing neutral gauge bosons other than $\Zp$ and $\Zpp$ in this model, 
but they only couple to new heavy fermions. 
So we discard their contributions by assuming that the new fermions are too heavy to be seen below a few TeV. 
 The neutral currents contributions involving $\Zp$ and $\Zpp$ are given by 
\beq \label{Eq:Ncurrents}
\mathcal{L}_{NC}^{\prime} =  
 - \frac{g}{2c_W}\sum_{\psi}\left[\bar{\psi}\gamma^\mu\left(g^{\prime}_V - g^{\prime}_A\gamma_5\right) \psi\right]Z^{\prime}_\mu 
 + \frac{g}{4\sqrt{2}}\sum_{\psi}\left[\bar{\psi}\gamma^\mu\left(1-\gamma_5\right) \psi\right]Z_\mu^{\prime\prime},
\eeq    
where $\psi$ are the SM fermions, 
and $g_V^{\prime}$ and $g_A^{\prime}$ are corresponding coupling constants 
of which values are listed in Table \ref{tab:couplings}.
\footnote{The contributions of the flavor mixings between the SM and the new heavy fermions 
to $g_V^{\prime}$ and $g_A^{\prime}$ are suppressed by $1/f^2$, so we neglect such flavor mixing effects.} 
The couplings in the table are family-universal while those in the K-S model with anomaly-free fermion embedding are not.
One can see from the table that the couplings contain additional new physics (NP) contributions proportional to the mixing angle $\theta$ between $Z$ and $Z'$ where $s_\theta \equiv \sin\theta =  t_W^2\sqrt{1-t_W^2}v^2/(2c_Wf^2)$.
Note that the masses of all heavy gauge bosons and the mixing angle $\theta$ are uniquely determined by the single parameter $f$. 
In the general case of $f_{12} \neq f_{34}$, 
we obtain the following condition:  $f_{12}^2 - f_{34}^2 = \left(v_1^2-v_2^2\right)\left(1+ O(v^2/f^2)\right) \ll f_{ij}^2$,
from the fact that there must be one zero eigenvalue corresponding to the photon.
Therefore, $f_{12}$ could not be much different from $f_{34}$, 
and our simplifying assumption is legitimate as a first approximation.

\begin{table}[hbt!]
\renewcommand{\arraystretch}{1.5}
\begin{tabular}{|c|c|c|}
   \hline 
  $\psi$ &  $g_V^\prime$ & $g_A^\prime$ \\
  \hline
  $t$ & $\left(\frac{1}{2}-\frac{4}{3}s_W^2\right)s_\theta - \frac{5}{6}rs_W$
      & $\frac{1}{2}s_\theta + \frac{1}{2}rs_W $ \\
  $b$ & $\left(-\frac{1}{2}+\frac{2}{3}s_W^2\right)s_\theta +\frac{1}{6}rs_W$
      & $-\frac{1}{2}s_\theta - \frac{1}{2}rs_W $ \\
$\nu$ & $\frac{1}{2}s_\theta +\frac{1}{2}rs_W$
      & $\frac{1}{2}s_\theta + \frac{1}{2}rs_W $ \\
  $e$ & $\left(-\frac{1}{2} + 2s_W^2\right)s_\theta +\frac{3}{2}rs_W$
      & $-\frac{1}{2}s_\theta - \frac{1}{2}rs_W $  \\   
  \hline  
\end{tabular}
\caption{$\Zp$ couplings to the SM fermions where $r \equiv g_X/g$.}
\label{tab:couplings}
\end{table}

 Instead of the little Higgs mechanism,  one can achieve the symmetry breaking for fermion and gauge boson masses 
by introducing the four SM type Higgs scalar fields with VEVs aligned as \cite{Ponce07}
\beqr \label{Eq:Higgsvev}
\langle \phi_1^T \rangle &=&  (v_u, 0, 0, 0) \sim [1, 4, -1/2],  \cr
\langle \phi_2^T \rangle &=&  (0, 0, V_u, 0) \sim [1, 4, -1/2],  \cr
\langle \phi_3^T \rangle &=&  (0, v_d, 0, 0) \sim [1, 4, 1/2],  \cr
\langle \phi_4^T \rangle &=&  (0, 0, 0, V_d) \sim [1, 4, 1/2],  
\eeqr  
where $V_u(V_d)$ corresponds to $f_{12}(f_{34})$ in the LHM.  With this choice of scalar sector, the theory becomes renormalizable
but the Higgs mass fine-tuning issue by the one-loop quadratic divergences remains intact in the model.
The symmetry breaking pattern of this model with the hierarchy $V_u \sim V_d \gg v_u, v_d$ is similar to that of the LHM,
due to the same gauge structure.
After EWSB with the VEVs in Eq. (\ref{Eq:Higgsvev}), 
three physical neutral gauge bosons $Z$, $\Zp$, and $\Zpp$ have the following masses (squared) similarly to Eq. (\ref{Eq:Zmass}):
\beq \label{Eq:Zmass2}
M^2_Z = \frac{g^2v_4^2}{c_W^2}\left(1 - t_W^4\frac{v_4^2}{V_4^2}\right), \quad
M^2_{\Zp} = (g^2+g_X^2)V_4^2 - r^2s_W^2M^2_{Z}, \quad
M^2_{\Zpp} = \frac{1}{2}g^2(V_4^2 + v_4^2) ,
\eeq 
where we use the following simplifying assumption $V_u = V_d \equiv V_4$ and $v_u = v_d \equiv v_4$ as done in Ref. \cite{Ponce07}. 
The $\Zpp$ mass can be expressed in terms of the lighter $Z$ and $\Zpp$ masses as
\beq \label{Eq:ZprimeMass}
M^2_{\Zp} = 2(1 + r^2)M^2_{\Zpp} - M^2_Z.
\eeq 
One can clearly see from Eq. (\ref{Eq:Zmass}) that the above mass relationship holds in the LHM with the same gauge group.
Once we introduce the same anomaly-free set of fermions given in Eq. (\ref{Eq:fermion}), 
the gauge boson couplings to the SM fermions are also similar to those in the LHM 
so that the low energy phenomenology of $\Zp$ and $\Zpp$ shall appear very similarly in both models.  
Therefore, the bounds of $\Zp$ and $\Zpp$ masses obtained in the LHM can be applicable to the model with regular Higgs mechanism.
Since there are additional scalars in the LHM comparing to the model with regular Higgs mechanism, 
one can discriminate the exotic neutral currents from those different models by investigating the scalar-gauge interactions.
But we postpone such study until the observation of the exotic neutral currents. 
Instead, we focus on the mass bounds of those exotic bosons, 
and consider their discovery potentials at the next run of the LHC in this paper.

\section{Decays of $\Zp$ and $\Zpp$}

 Since the neutral current interactions are identified, we can now proceed to analyze the $\Zp$ and $\Zpp$ decays and production rates.
Other than the neutral currents discussed in the previous session, 
$WW\Zp$ triple gauge boson coupling and $HZ\Zp$ Higgs gauge boson coupling arise in this model through $Z-\Zp$ mixing
so that $\Zp$ can decay into $WW$ and $ZH$ final states while $\Zpp$ cannot.
From the neutral currents given in Eq. (\ref{Eq:Ncurrents}),
we obtain the $\Zp$ decay rates as
\beqr \label{Eq:Decayrates}
\Gamma(\Zp \to \psi\bar{\psi})   
  &=& N_c\frac{G_F}{6\sqrt{2}\pi} M^2_Z M_{\Zp} \left[\left(\gVps + \gAps \right)\left(1 - \dfrac{M_\psi^2}{M^2_{\Zp}}\right) 
    + 3\left(\gVps - \gAps \right)\dfrac{M_\psi^2}{M^2_{\Zp}}  \right] \left(1 - 4\dfrac{M_\psi^2}{M^2_{\Zp}}\right)^{1/2}, \cr 
\Gamma(\Zp \to W^+ W^-)   
  &=&   \frac{G_F}{24\sqrt{2}\pi} c_W^4 s_\theta^2  M^2_Z M_{\Zp} 
  \left( \frac{M^4_{\Zp}}{M^4_W} + 20\frac{M^2_{\Zp}}{M^2_W} + 12 \right) \left(1 - 4\dfrac{M_W^2}{M^2_{\Zp}}\right)^{3/2}, \cr
\Gamma(\Zp \to Z H)   
  &=&   \frac{G_F}{6\sqrt{2}\pi}  s_\theta^2  M^2_Z M_{\Zp} \left[ 2\dfrac{M_Z^2}{M^2_{\Zp}} 
    + \dfrac{1}{4}\left(1 + \dfrac{M_Z^2}{M^2_{\Zp}} - \dfrac{M_H^2}{M^2_{\Zp}}  \right)^2\right] \cr
  && \times   \left[ 1 -\left (\dfrac{M_Z^2}{M^2_{\Zp}} + \dfrac{M_H^2}{M^2_{\Zp}}\right)^2  \right]^{1/2}
     \left[ 1 - \left(\dfrac{M_Z^2}{M^2_{\Zp}} - \dfrac{M_H^2}{M^2_{\Zp}}\right)^2  \right]^{1/2} .
\eeqr
The $\Zpp$ decay rates into fermion pairs are given similarly to the $\Zp$ decay rates
by replacing the couplings $g_V^{\prime}$ and $g_A^{\prime}$ in Eq. (\ref{Eq:Decayrates})
with the flavor universal $\Zpp$ coupling 
to the SM fermions given in Eq. (\ref{Eq:Ncurrents}).  
The new fermion masses are comparable to the new gauge boson masses. 
In this work, we assume that those new fermions such as $T$ and $B$ are too heavy  
for the neutral gauge bosons to decay into them \cite{Nam10}.  
The decay rates of $\Zp \to W^+ W^-$ and $\Zp \to Z H$ are suppressed 
by $Z-\Zp$ mixing angle squared $s_\theta^2 \sim (v/f)^4$.
However, the triple gauge boson coupling is proportional to the incoming and outgoing momentum, 
and the suppression by $Z-\Zp$ mixing is compensated by the heavy $\Zp$ mass contribution $(M_{\Zp}/M_W)^4  \sim (f/v)^4$
so that the decay rate of $\Zp \to W^+ W^-$ is comparable to those of $\Zp \to \psi\bar{\psi}$ 
as shown in Eq. (\ref{Eq:Decayrates}).
On the other hand, there is no enhancement in the decay of $\Zp \to Z H$ 
so that the corresponding rate is negligible in this model. 

\begin{figure}[!hbt]
\centering%
  \subfigure[ ]{\label{Branchingf1} %
    \includegraphics[width=7.5cm]{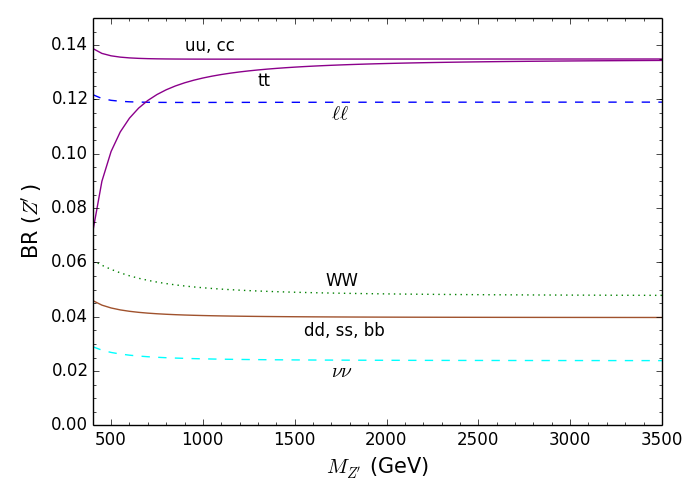}} \qquad
  \subfigure[ ]{\label{Branchingf2} %
    \includegraphics[width=7.5cm]{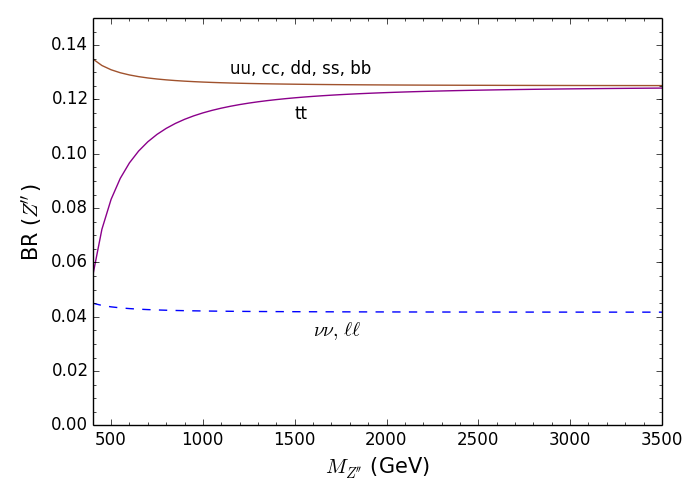}}
\caption{Branching ratios of (a) $\Zp$ boson and of (b) $\Zpp$ boson as a function of their masses.} 
\label{Branchingf}
\end{figure}

\begin{figure}[!hbt]
\centering%
  \subfigure[ ]{\label{Decayrate1} %
    \includegraphics[width=7.5cm]{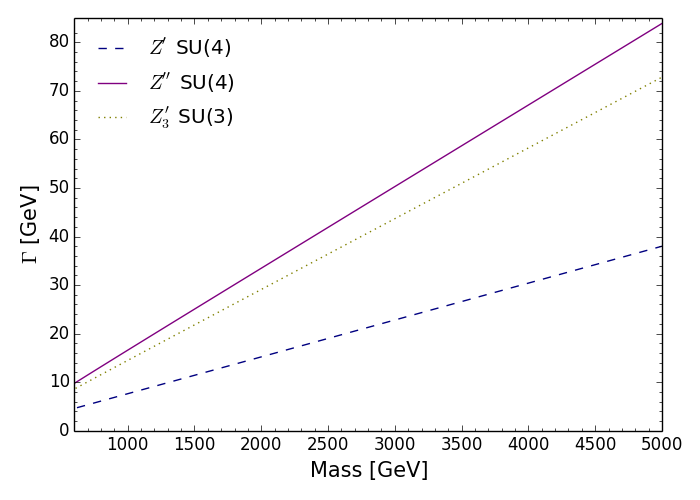}} \qquad
  \subfigure[ ]{\label{Decayrate2} %
    \includegraphics[width=7.5cm]{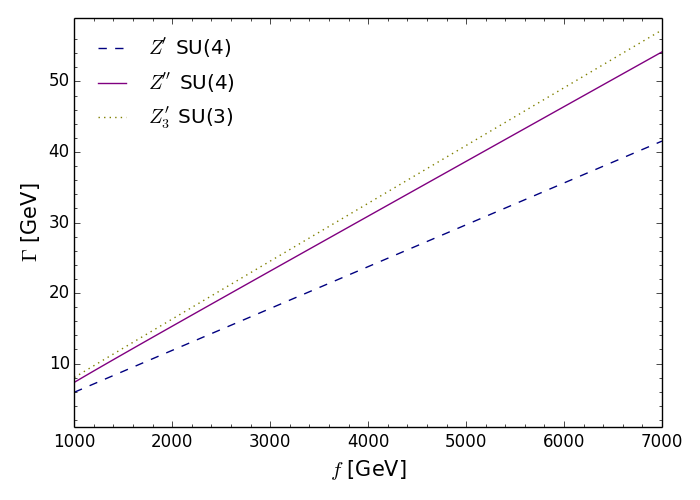}}
\caption{Decay widths of $\Zp$ and $\Zpp$ bosons as a function of (a) their masses and of (b) the scale parameter $f$.
$Z_3^\prime$ denotes the extra neutral gauge boson in the $SU(3)_L \times U(1)_X$ LHM.} 
\label{Decayrate}
\end{figure}

 We illustrate the branching ratios of $\Zp$ and $\Zpp$ bosons in Fig. \ref{Branchingf} 
as a function of $M_{\Zp}$ and $M_{\Zpp}$, 
which show that the decay patterns of  $\Zp$ and $\Zpp$ are obviously different.  
In Fig. \ref{Decayrate}, we show the total decay widths of $\Zp$ and $\Zpp$ bosons 
as a function of their masses and of the scale parameter $f$.  
In the figure, as a comparison, we also show the total decay width of the extra neutral gauge boson 
denoted as $Z_3^\prime$ which appears in the $SU(3)_L \times U(1)_X$ LHM \cite{Schmaltz04, Han06}.
If $\Zp$ or $\Zpp$ decays into new heavy fermion pairs,
the slope of each curve in the Fig. \ref{Decayrate} increases from the resonance masses of the new fermions.  
But we assume $M_{\Zp}, M_{\Zpp} < 2M_F$ for all new fermions $F$,
and Fig.  \ref{Decayrate} just shows uniform slopes. 
Taking consideration of all the possibilities for the new fermion masses is beyond the scope of this paper, 
and our results in the next sections are obtained with no new heavy fermionic resonances.
Even if $M_{\Zp}, M_{\Zpp} > 2M_F$ for some of new fermions $F$, still, 
branching ratios of $\Zp$ and $\Zpp$ decaying into the new fermions would be only a few percent 
due to the small fermion mixing. 
Hence the decreases of the $\Zp$ and $\Zpp$ cross-section times the SM fermion branching ratios should be very small \cite{Coutinho13}.
Therefore, the bounds obtained in the selected channels discussed here shall not change much.

\section{Current bounds at $\sqrt{s} = 8$ TeV}

 In order to generate processes, we implemented the LHMs in the MadGraph5 package \cite{MadGraph5} 
using Feynrules \cite{Feynrules}.  
For cross-section calculation and event generation, we used Pythia \cite{Pythia6}, 
and simulated the signal events of four different processes, 
dilepton, $\tau^+\tau^-$, dijet, and $t\bar{t}$, at the CM energy of 8 TeV.
The $\Zp$ and $\Zpp$ cross-section times branching ratios $(\sigma\times BR)$ are calculated 
within a range of $\pm3\Gamma$ 
around the $\Zp$ and $\Zpp$ pole masses as done similarly in Ref. \cite{Dittmar04}. 
For experimental bounds, we used the 8 TeV collision data collected by the CMS and the ATLAS experiments. 
The combined dilepton data by the CMS group correspond to an integrated luminosity of 19.6 fb$^{-1}$ in the dielectron channel
and 20.6 fb$^{-1}$ in the dimuon channel \cite{CMS13lepton0}, 
and those by the ATLAS group correspond to an integrated luminosity of 20.3 fb$^{-1}$ in the dielectron channel
and 20.5 fb$^{-1}$ in the dimuon channel  \cite{ATLAS13lepton0}.
The dijet data collected by the CMS and the ATLAS groups correspond to an integrated luminosity of 19.6 fb$^{-1}$ \cite{CMS13jet0}
and 13 fb$^{-1}$ \cite{ATLAS12jet0}, respectively.
The most recent ditau and ditop resonance searches have been performed by the ATLAS group 
using 19.5 fb$^{-1}$ data in the fully hadronic channel \cite{ATLAS13tau0} 
and 14 fb$^{-1}$ data in the lepton plus jets channel \cite{ATLAS13top0}, respectively. 

\begin{figure}[!hbt]
\centering%
  \subfigure[ ]{\label{Boundsll} %
    \includegraphics[width=7.5cm]{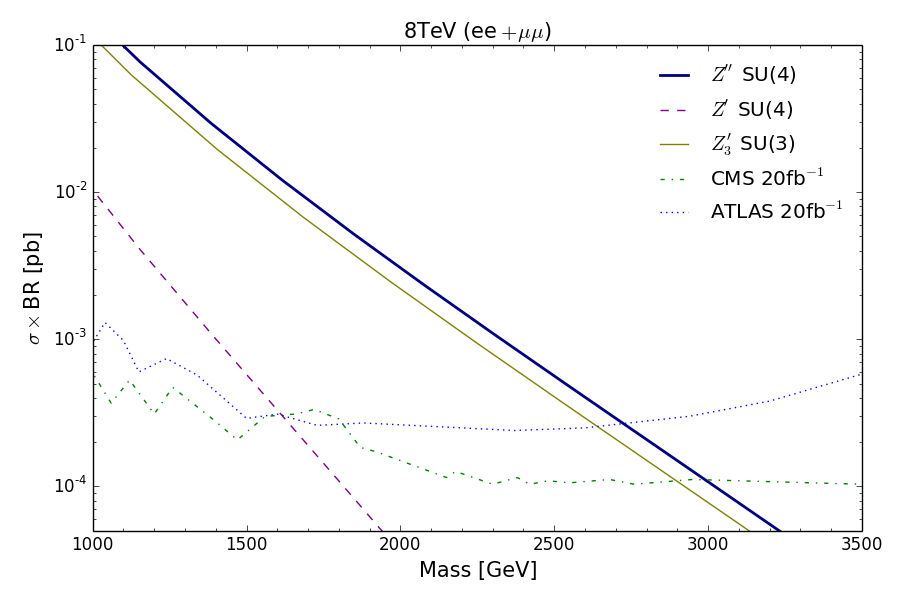}} \qquad
  \subfigure[ ]{\label{Boundstau} %
    \includegraphics[width=7.5cm]{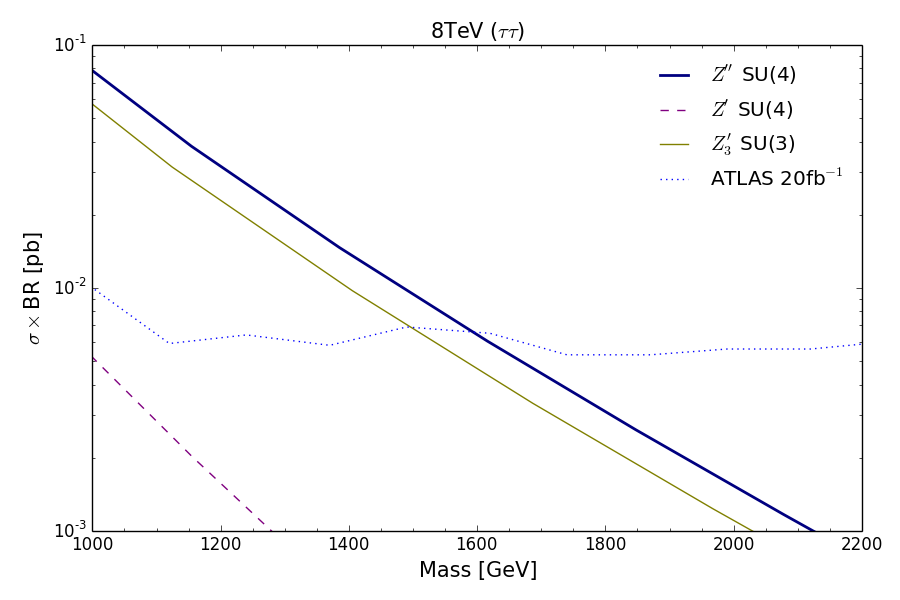}} \\
  \subfigure[ ]{\label{Boundsjj} %
    \includegraphics[width=7.5cm]{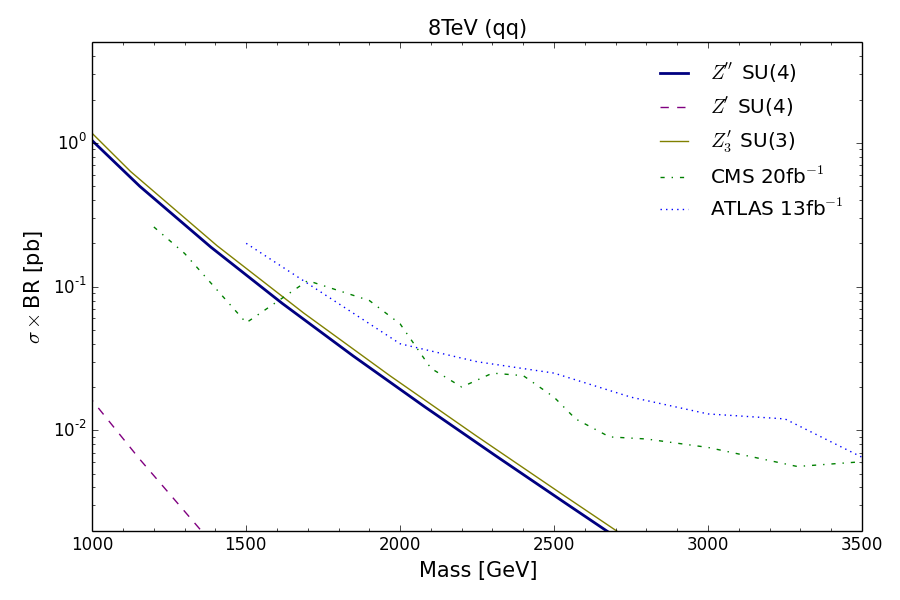}} \qquad
  \subfigure[ ]{\label{Boundstt} %
    \includegraphics[width=7.5cm]{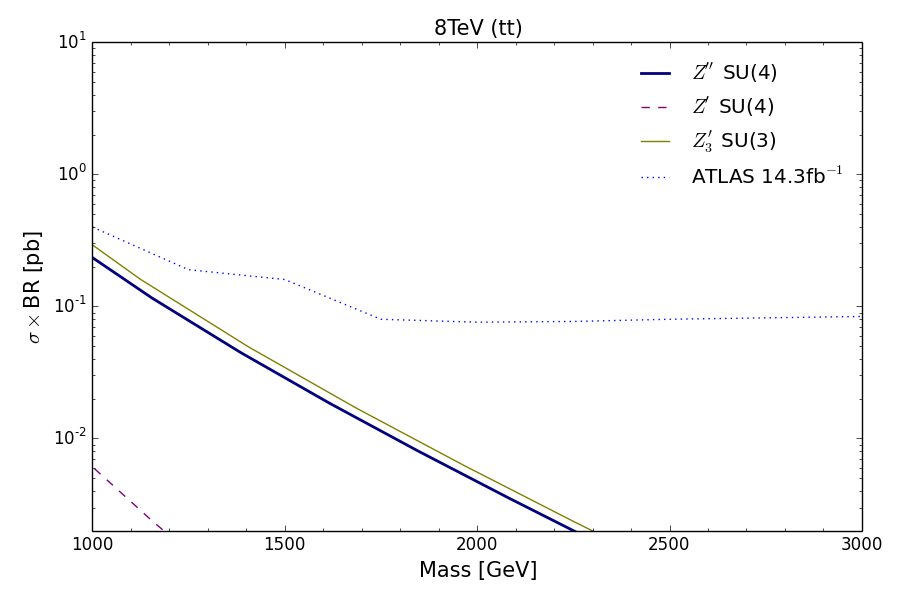}}  
\caption{Observed upper cross-section times branching ratio $(\sigma\times BR)$ limits at 95$\%$ CL
for $\Zp$ and $\Zpp$ bosons in the $SU(4)_L \times U(1)_X$ LHM and for $\Zp_3$ bosons in the $SU(3)_L \times U(1)_X$ LHM
using (a) combined dilepton, (b) $\tau^+\tau^-$, (c) dijet, and (d) $t\bar{t}$ channels.} 
\label{Bounds}
\end{figure}

 In Fig. \ref{Bounds}, we show the observed exclusion limits on the cross-section times branching ratio at 95$\%$ CL
for $\Zp$ and $\Zpp$ bosons in the $SU(4)_L \times U(1)_X$ LHM and for $\Zp_3$ bosons in the $SU(3)_L \times U(1)_X$ LHM
for the dilepton, $\tau^+\tau^-$, dijet, and $t\bar{t}$ channels.  
Note in the figures that $\sigma\times BR$ for $\Zp$ are depicted as a function of the $\Zpp$ mass
since the $\Zp$ and $\Zpp$ masses are dependent each other and determined by a single parameter $f$. 
One can see from the figures that the dilepton measurements give the strongest bounds on the $\Zp$ and $\Zpp$ boson masses
as well as on the $\Zp_3$ mass.
In the dilepton channel, the CMS limit in Ref. \cite{CMS13lepton0} was given on the production ratio $R_\sigma$ of cross-section times branching fraction 
for $\Zp$ bosons to the same quantity for $Z$ bosons, 
but we converted it to $\sigma\times BR$ using the $Z$ cross-section obtained within the range of 60$\textendash$120 GeV \cite{CMS14boson},
for a clear comparison with the other measurements.
As shown in the figures, $\Zpp$ in the $SU(4)_L \times U(1)_X$ model is excluded below 2980 GeV by the CMS measurements
and 2730 GeV by the ATLAS measurements. 
Note that the corresponding ATLAS result has been recently published \cite{ATLAS13lepton0} 
while the CMS result is not \cite{CMS13lepton0}, as of now.
The bounds on the $\Zp_3$ mass obtained from the dilepton measurements are about 100 GeV lower than those of $\Zpp$. 
Using the mass relationship given in Eq. (\ref{Eq:ZprimeMass}), 
the lower mass bound of $\Zp$ boson is also obtained as 5040 (4620) GeV with respect to the CMS (ATLAS) dilepton measurement.   
For reference, if we use the best previous published results of up to 5 fb$^{-1}$ data collected 
at $\sqrt{s} = 7$ and 8 TeV \cite{CMS13lepton, ATLAS12lepton}, we obtain the lower mass limit of $\Zpp$ to be 2500 GeV
which corresponds to the $\Zp$ mass of 4230 GeV.
Instead of the arbitrarily chosen $3\Gamma$ interval used in the present analysis for cross-section calculations,
if we choose the events in a range of 40$\%$ of the $\Zpp$ mass as done in Ref. \cite{CMS13lepton0},  
the mass bound rises about 70 GeV.

\section{Discovery potentials at $\sqrt{s} = 14$ TeV}

\begin{figure}[!hbt]
\centering%
    \includegraphics[width=8cm]{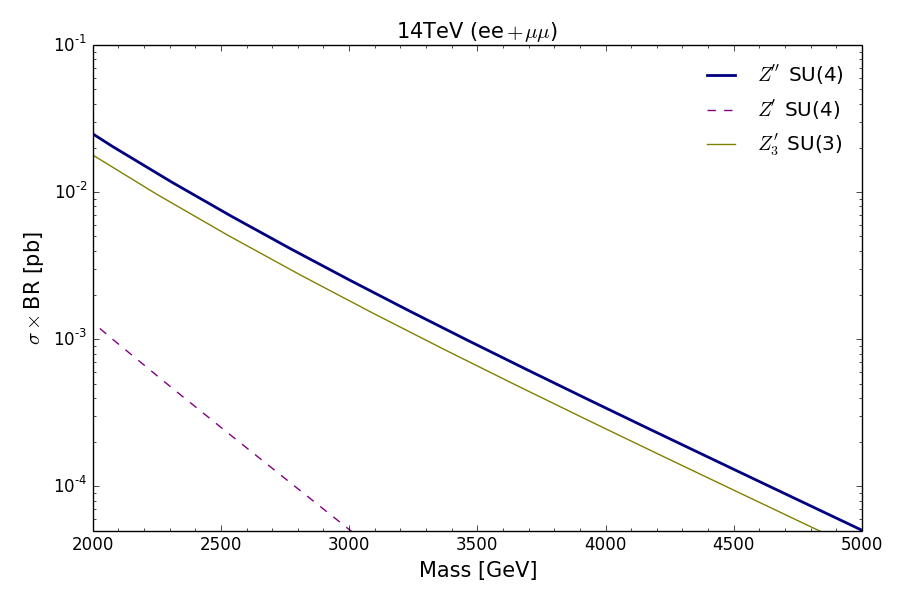}
\caption{Total cross-sections for the processes $p  p \to \Zpp(\Zp, \Zp_3) \to \ell^+ \ell^-$ 
as a function of the $\Zpp$ and $\Zp_3$ masses at $\sqrt{s}=$14 TeV.
The plot for $\Zp$ is depicted as a function of the $\Zpp$ mass.} 
\label{CS14}
\end{figure}

 After two years of intense maintenance, the LHC is back in operation at a CM energy of 13 TeV in April 2015, 
and will reach at the designed energy of 14 TeV in a few years. 
In this section, we investigate the LHC potential to find a $\Zpp$ as well as $\Zp_3$ at 14 TeV,
especially in the dilepton channel.
Fig. \ref{CS14} shows the total cross-sections calculated at tree level for the processes 
$p  p \to \Zpp(\Zp, \Zp_3) \to \ell^+ \ell^-$ at 14 TeV, with $\ell = e,\mu$ .  
One can see from the figures that depending on the masses of $\Zpp(\Zp, \Zp_3)$, 
the cross-sections increase by a factor of 10 to $10^2$ at 14 TeV in comparison with their values at 8 TeV shown in Fig. \ref{Boundsll}.
 
\begin{figure}[!hbt]
\centering%
    \includegraphics[width=9cm]{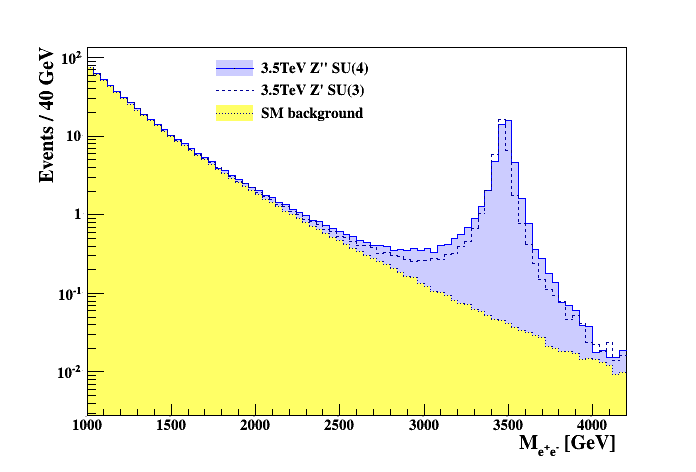}
\caption{Invariant mass distributions in the electron channel for the new exotic neutral gauge bosons with their masses set to 3.5 TeV   
using a luminosity of 100 fb$^{-1}$ data at $\sqrt{s}=$14 TeV.} 
\label{Invmass}
\end{figure}

 The decays $\Zpp(\Zp, \Zp_3) \to \ell^+ \ell^-$ provide a clean signature in the mass spectrum. 
The dominant background in this case is the irreducible Drell-Yan (DY) process. 
The contribution of the other backgrounds is less than 30$\%$ of the DY cross-section
and can be heavily suppressed by isolation cuts at high masses, so we consider only the DY background \cite{Pedraza09}.
Since the interference between the $\Zpp(\Zp, \Zp_3)$ and the $Z/\gamma^\ast$ processes is minimal,    
we also treat signals independent of backgrounds.  
In Fig. \ref{Invmass}, we shows the invariant distribution for the dielectron system,
as expected for the $SU(4)_L \times U(1)_X$ LHM with $M_{\Zpp} = 3.5$ TeV 
and for the SM background using a luminosity of 100 fb$^{-1}$ data at 14 TeV. 
We also show the case for the $SU(3)_L \times U(1)_X$ LHM with $M_{\Zp_3} = 3.5$ TeV as a comparison.
One can see from the figure that the width of $\Zp_3$ boson is narrower than that of $\Zpp$ as also shown in Fig. \ref{Decayrate1}.
Although only the invariant mass distribution for the electron channel is shown in Fig. \ref{Invmass}, 
the distribution for the muon channel looks nearly the same. 

\begin{figure}[!hbt]
\centering%
    \includegraphics[width=8cm]{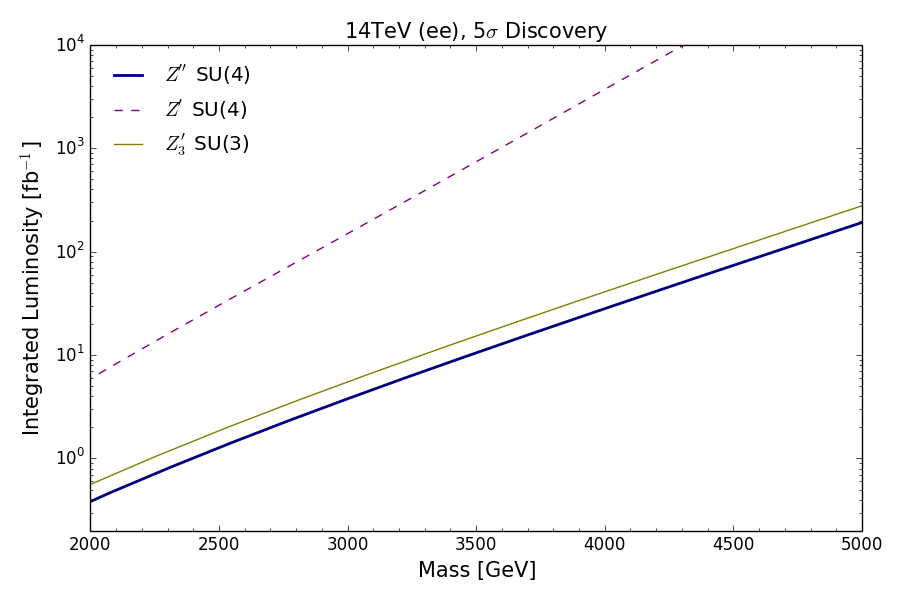}
\caption{Integrated luminosity needed for 5$\sigma$ discovery of $\Zpp(\Zp, \Zp_3) \to e^+e^-$
as a function of the $\Zpp$ and $\Zp_3$ masses at $\sqrt{s}=$14 TeV.
The plot for $\Zp$ is depicted as a function of the $\Zpp$ mass.} 
\label{Discovery}
\end{figure}

 The discovery potential to find a new exotic particle is determined by the integrated luminosity needed for a 5$\sigma$ excess.
The statistical significance is obtained from the expected number of signal($N_s$) and background($N_b$) events
within a chosen exotic particle mass window ($3\Gamma$ in our study), and defined as \cite{Cowan11}:
\beq
S = \sqrt{2\left((N_s + N_b)\textrm{ln}(1 + N_s/N_b) -N_s\right)} ,
\eeq
which gives a good approximation to the likelihood ratio based significance in the low statistical regime.
For more realistic study, we consider an overall efficiency of 73$\%$ for the electron channel 
as determined by the ATLAS experiment \cite{ATLAS13lepton0}.
In Fig. \ref{Discovery}, we shows the expected integrated luminosity needed for 5$\sigma$ discovery in the electron channel
as a function of the $\Zpp$ mass for the $SU(4)_L \times U(1)_X$ LHM 
and of the $\Zp_3$ mass for the $SU(3)_L \times U(1)_X$ LHM.   
The figure shows that, even with a total integrated luminosity of a few fb$^{-1}$, 
it would be possible to reach the limits set by previous searches of this type of neutral gauge boson.  
For $M_{\Zpp} \sim 3.5$ TeV in the $SU(4)_L \times U(1)_X$ LHM, it is required to have about 12 fb$^{-1}$ data
to discover this new heavy state, 
while about 5 fb$^{-1}$ more data are needed to observe $\Zp_3$ boson expected in the $SU(3)_L \times U(1)_X$ LHM.  
It is not trivial to identify the origin of the observed neutral bosons in the discussed decay channels until 100 fb$^{-1}$ data are collected
above the obtained mass bound.
As also shown in the figure, if we observe a new neutral boson with a mass of 3.5 TeV for instance, 
we need to collect about $10^3$ fb$^{-1}$ data to see if there is another exotic boson 
in order to discriminate the models from a direct observation. 
Otherwise, we should take into account to investigate other physical observables as well such as the forward-backward charge asymmetry 
although it is non-trivial to measure such a quantity at the LHC because the original quark direction is unknown \cite{Dittmar04},
and we are not going that far in this paper.
As discussed in Sec. II, our obtained results are valid on the mass bounds and the LHC discovery potentials 
in the similar models with the standard Higgs mechanism as far as the gauge structures and the fermion types are the same.

\section{Conclusions}

 In this paper, we obtained the lower bounds on the masses of $\Zp$ and $\Zpp$ bosons appeared 
in the $SU(4)_L \times U(1)_X$ LHM with anomaly-free set of fermions using the present LHC data.     
In this model, $\Zpp$ boson is lighter than $\Zp$ boson, and their masses are determined by the single scale parameter $f$ 
with the relationship given in Eq. (\ref{Eq:ZprimeMass}). 
The strongest lower mass bound is obtained currently in the dilepton channel 
with about 20 fb$^{-1}$ data collected at $\sqrt{s} =$ 8 TeV.
The CMS result from the analysis of the production ratio $R_\sigma$ excludes $\Zpp$ boson with mass below 2980 GeV 
while the ATLAS result from the analysis of $\sigma\times BR$ excludes $\Zpp$ boson with mass below 2730 GeV. 
For comparison, we also considered $\Zp_3$ boson appeared in the $SU(3)_L \times U(1)_X$ LHM 
with anomaly-free set of fermions,
and the exclusion limit of the $\Zp_3$ mass is about 100 GeV lower than that of the $\Zpp$ mass.

 The search for these new neutral bosons is an important issue of the experimental program of the LHC 
running at the designed CM energy of 14 TeV.
In order to see the discovery potentials of $\Zp$ and $\Zpp$ bosons,
we presented in Fig. \ref{Discovery} the expected integrated luminosity needed for $5\sigma$ discovery
in the electron channel as a function of new exotic neutral boson masses.
For $M_{\Zpp} \sim 3.5$ TeV, it is required to have about 12 fb$^{-1}$ data to discover this new heavy state, 
while about 5 fb$^{-1}$ more data are needed to observe $\Zp_3$ boson with the same mass.
With a luminosity of 100 fb$^{-1}$ data, one can search for $\Zpp(\Zp_3)$ boson with mass up to about 4650(4450) GeV
in the electron channel.  Of course, if the detector efficiency is improved,  
the search region of the exotic boson masses can be extended further.   
 
 As discussed earlier, our results can be applicable to the $SU(4)_L \times U(1)_X$ model with the standard Higgs mechanism
as far as the fermion structure is the same.  
The different models could be discriminated either by observing the extra scalars appeared in the models
or by investigating the other properties of the exotic bosons such as charge asymmetry and rapidity distribution simultaneously.
This study shall be proceeded further as future experimental progress reveals a hint on a exotic state.

\begin{acknowledgments}
The numerical computation of this work was supported in part by PLSI resources of KISTI.
KYL is supported by the Basic Science Research Program through the National Research Foundation
of Korea (NRF) funded by the Korean Ministry of Education, Science and Technology (2010-0010916).
\end{acknowledgments}

\def\npb#1#2#3 {Nucl. Phys. B {\bf#1}, #2 (#3)}
\def\plb#1#2#3 {Phys. Lett. B {\bf#1}, #2 (#3)}
\def\prd#1#2#3 {Phys. Rev. D {\bf#1}, #2 (#3)}
\def\jhep#1#2#3 {J. High Energy Phys. {\bf#1}, #2 (#3)}
\def\jpg#1#2#3 {J. Phys. G {\bf#1}, #2 (#3)}
\def\epj#1#2#3 {Eur. Phys. J. C {\bf#1}, #2 (#3)}
\def\arnps#1#2#3 {Ann. Rev. Nucl. Part. Sci. {\bf#1}, #2 (#3)}
\def\ibid#1#2#3 {{\it ibid.} {\bf#1}, #2 (#3)}
\def\none#1#2#3 {{\bf#1}, #2 (#3)}
\def\mpla#1#2#3 {Mod. Phys. Lett. A {\bf#1}, #2 (#3)}
\def\pr#1#2#3 {Phys. Rep. {\bf#1}, #2 (#3)}
\def\prl#1#2#3 {Phys. Rev. Lett. {\bf#1}, #2 (#3)}
\def\ptp#1#2#3 {Prog. Theor. Phys. {\bf#1}, #2 (#3)}
\def\rmp#1#2#3 {Rev. Mod. Phys. {\bf#1}, #2 (#3)}
\def\zpc#1#2#3 {Z. Phys. C {\bf#1}, #2 (#3)}
\def\cpc#1#2#3 {Comput. Phys. Commun. {\bf#1}, #2 (#3)}

\end{document}